\documentclass[useAMS,usenatbib]{mn2e}
\usepackage{aas_macros,graphicx,times,multirow}

\newcommand{\fermi}{{\sl Fermi}}

\newcommand{\latn}{{Large Area Telescope}}
\newcommand{\lat}{{\sl LAT}}
\newcommand{\rosn}{{\sl R\"ontgen Satellite}}
\newcommand{\ros}{{\sl ROSAT}}
\newcommand{\asca}{{\sl ASCA}}
\newcommand{\ascan}{{\sl Advanced Satellite for Cosmology and Astrophysics}}
\newcommand{\chan}{{\sl Chandra}}
\newcommand{\xmm}{{\sl XMM-Newton}}
\newcommand{\vltn}{{\rm Very Large Telescope}}
\newcommand{\vlt}{{\rm VLT}}
\newcommand{\fors}{{\rm FORS2}}

\newcommand{\gron}{{\sl Compton Gamma Ray Observatory}}
\newcommand{\gro}{{\sl CGRO}}

\def \psr{PSR\, J1048$-$5832}

\title[New VLT observations of the $Fermi$ pulsar PSR\, J1048$-$5832]{New VLT observations of the $Fermi$ pulsar PSR\, J1048$-$5832}
\author[M. Razzano, R. P. Mignani, M. Marelli, A. de Luca. ]{M. Razzano$^{1,2,3}$, R.P. Mignani$^{4,5}$, M. Marelli$^{6,7}$, A. de Luca$^{6,8}$\footnotemark[1]\thanks{Based on observations made with ESO Telescopes at the La Silla Paranal Observatory under programme ID 384.D-0386(A), 386.D-0585(A).}\\
$^{1}$ Istituto Nazionale di Fisica Nucleare, Sezione di Pisa, I-56127 Pisa, Italy \\
$^{2}$ Dipartimento di Fisica ``E. Fermi'', Universit\`a di Pisa, 56127 Pisa, Italy \\
$^{3}$ Santa Cruz Institute for Particle Physics, University of California, Santa Cruz, CA 95064 \\
$^{4}$ Mullard Space Science Laboratory, University College London, Holmbury St. Mary, Dorking, Surrey, RH5 6NT, UK\\
$^{5}$ Kepler Institute of Astronomy, University of Zielona G\'ora, Lubuska 2, 65-265, Zielona G\'ora, Poland \\
$^{6}$ INAF - Istituto di Astrofisica Spaziale e Fisica Cosmica Milano, via E. Bassini 15, 20133, Milano, Italy\\
$^{7}$ Universit\'a degli Studi dell' Insubria, Via Ravasi 2, 21100, Varese, Italy \\
$^{8}$ INFN - Istituto Nazionale di Fisica Nucleare, sezione di Pavia, via A. Bassi 6, 27100, Pavia, Italy
}
\begin{document}

\date{Accepted 1988 December 15. Received 1988 December 14; in original form 1988 October 11}

\pagerange{\pageref{firstpage}--\pageref{lastpage}} \pubyear{2002}

\maketitle

\label{firstpage}

\begin{abstract}
 PSR\, J1048$-$5832 is a Vela-like (P=123.6 ms; $\tau\sim$20.3 kyr)  $\gamma$-ray pulsar detected by \fermi, at a distance of $\sim$2.7 kpc and with a rotational energy loss rate  $\dot{E}_{SD}  \sim 2 \times 10^{36}$ erg s$^{-1}$.  The \psr\ field has been observed with the  \vltn\ (\vlt)  in the $V$ and $R$ bands. We used these data to determine the colour of the object detected closest to the  \chan\ position  (Star D) and confirm that it is not associated with the pulsar.  For the estimated extinction along the line of sight, inferred from a re-analysis of the \chan\  and \xmm\ spectra, the fluxes of Star D ($V\sim26.7$; $R\sim25.8$) imply a $-0.13 \la (V-R)_0 \la 0.6$. This means that the \psr\ spectrum would be unusually red compared to the Vela pulsar.
 Moreover, the ratio between the unabsorbed optical and X-ray flux of \psr\ would be much higher than for other young pulsars.  Thus, we conclude that Star D is not the \psr\  counterpart. We compared the derived R and V-band upper limits ($R\ga$26.4; $V\ga$27.6) with the extrapolation of the X and $\gamma$-ray spectra and constrained the pulsar spectrum at low-energies. In particular, the VLT upper limits suggest that the pulsar  spectrum could be consistent with a single power-law, stretching from the $\gamma$-rays to the optical.
\end{abstract}

\begin{keywords}
Optical: stars -- neutron stars
\end{keywords}

\section{Introduction}

PSR\, J1048$-$5832 (B1046$-$58) is a young, Vela-like pulsar in the Carina region, discovered during a 1.4 GHz Parkes survey of the Galactic plane (Johnston et al.\ 1992). It has a period $P$=123.7 ms and a period derivative $\dot{P}$=9.63 $\times$ 10$^{-14}$ s s$^{-1}$, which correspond to a characteristic age $\tau_{c}\sim$20.3 kyr, a surface dipole magnetic field $B_{S}\sim$3.5 $\times$ 10$^{12}$ G, and a spin down luminosity $\dot{E}_{SD}\sim$2$\times$10$^{36}$ erg s$^{-1}$. The NE2001 model (Cordes \& Lazio 2002) yields a distance of 2.7$\pm$0.35 kpc, based on the dispersion measure (DM). In the X-rays, PSR\, J1048$-$5832 was detected with the \rosn\ (\ros) at 0.1--2.4 keV (Becker et al.\ 1997). Soon after, observations with  the \ascan\ (\asca)  found possible evidence of extended X-ray emission (Pivoravoff et al.\ 2000), possibly associated with a pulsar wind nebula (PWN).  High-resolution observations with the \chan\ X-ray Observatory, indeed revealed the presence of an asymmetric PWN around the pulsar. No X-ray pulsed emission has been detected so far by both \chan\ and \xmm, and a 3$\sigma$ upper limit of 53$\%$ on the pulsed fraction  between 0.5 and 10 keV has been determined (Gonzales et al.\ 2006). A recent re-analysis of the \chan\  data of \psr\ (Marelli 2012) could separate the pulsar emission from that of its PWN. See Marelli (2012) for details on the observations and data analysis. The combined \chan\ plus \xmm\ spectrum of the pulsar is fitted by a power-law (PL) with photon index $\Gamma_{\rm X}$=1.35$\pm$0.45 and  $N_{\rm H}$=$(46.0\pm2.3) \times 10^{20}$ cm$^{-2}$, corresponding to an unabsorbed X-ray flux $F_{\rm X}$=$0.490^{+0.181}_{-0.342} \times 10^{-13}$ erg cm$^{-2}$ s$^{-1}$ in the 0.3--10 keV energy range. We note that, using the \chan\ data, Gonzales et al.\  (2006) tried to separate the pulsar and PWN  components and derived a pulsar photon index $\Gamma_{\rm X} \sim 2.4$ (quoted without uncertainties) from the fit to the combined \chan\ plus \xmm\ spectrum.
Our results differ from theirs since we used data from all the \xmm\ detectors (both the EPIC-PN and MOS) and fitted the combined spectrum using the C-statistic approach in {\sc XSPEC}. This allowed us to more efficiently separate the pulsar and PWN components in our spectral analysis and derive more accurate values for the pulsar photon index. Moreover, we used slightly different radii to extract the pulsar counts:  20\arcsec\  instead of 25\arcsec\ for the \xmm\ data, and  2\arcsec\  instead of 1\arcsec\ for the \chan\ data.
 PSR\, J1048$-$5832 was initially proposed by Kaspi et al.\ (2000) as the counterpart of the unidentified $\gamma$-ray source 3EG\, J1048$-$5840, detected above 400 MeV  by the EGRET instrument aboard the \gron\ (\gro). More recently, the \latn\ (\lat) aboard the {\em Fermi} Gamma-ray Space Telescope detected for the first time $\gamma$-ray pulsations above 100 MeV from this pulsar (Abdo et al.\ 2009), highlighting a clear, double-peaked pulse profile very similar to the Vela pulsar. Deep optical observations  of \psr\  were performed with the \vltn\ (\vlt) but  only one object (Star D; $V\sim$26.7) was detected close to the pulsar \chan\ position (Mignani et al.\ 2011).  Here, we present new \vlt\ observations of the \psr\ field in the R band that, together with the revised value of the extinction along the line of sight obtained from our re-analysis of the \chan\ and \xmm\ data,  confirm that  Star D cannot be the pulsar optical counterpart.

\section{Observations and data reduction}

The R-band images of \psr\  were obtained with the \vlt\ Antu telescope at the ESO Paranal Observatory between December 4, 2010 and January 5, 2011 (See Tab.\ 1) and are available in the ESO archive\footnote{www.eso.org/archive} under programme ID 384.D-0386(A) and 386.D-0585(A). All observations were performed in service mode with the FOcal Reducer/low dispersion Spectrograph (FORS2, Appenzeller et al.\ 1998). FORS2 is equipped with a red-sensitive MIT detector, a mosaic of two 2k$\times$4k CCDs optimised for wavelengths longer than 6000 \AA. The pixel size of FORS2 in standard resolution mode is 0\farcs25 ($2 \times 2$ binning), corresponding to a field of view of $8\farcm3 \times 8\farcm3$ over the CCD mosaic. 
Observations were performed in IMAGE mode, with standard low gain, normal readout (200 Kpix/s) and standard-resolution mode. For all the observations, the target was positioned on CHIP1. The R$_{\rm SPEC}$ filter ($\lambda$ = 6550 \AA; $\Delta \lambda$ = 1650 \AA) was used for all observations. In order to minimise the saturation of bright stars and to allow for cosmic ray removal, a series of 23 short exposures of 599.9 s were obtained, for a total exposure of 13797.7 s (Tab.\ 1). Exposures were taken in dark time and photometric conditions. The average airmass was below 1.5 and the seeing better than 1\arcsec, as 
determined by the  {\em Differential Image Motion Monitor} (DIMM)\footnote{archive.eso.org/asm/ambient-server} 

   \begin{table}
      \caption[]{Summary of the VLT R-band observations of \psr\ including: the exposure times (T), the number of exposures (N), the airmass sec(z) averaged over the N exposures, the seeing determined by the DIMM and its associated rms in (parentheses). }
         \label{Table1}
     $$ 
         \begin{array}{p{0.5\linewidth}cccc}   \hline
            \noalign{\smallskip}
            Date      &  T^{\mathrm{a}}  & N & sec($z$) & seeing\\
            YYYY-MM-DD      &  (s)  &  &  & (\arcsec)\\
            \noalign{\smallskip}
            \hline
            \noalign{\smallskip}
            2010-12-04      &  599.9 & 5 & 1.40 & 0.80 (0.02)\\
	        2011-01-01      &  599.9 & 8 & 1.47 & 0.86 (1.17)\\
	        2011-01-04      &  599.9 & 5 & 1.41 & 0.91 (0.09)\\
	        2011-01-05      &  599.9 & 5 & 1.25 & 0.77 (0.12)\\
            \noalign{\smallskip}
            \hline
         \end{array}
     $$ 
   \end{table}
   
We used the standard packages in {\sc IRAF} for the bias subtraction and for flat field correction, selecting the closest-in-time bias and twilight flat field images available on the ESO archive. After these standard corrections, we aligned and stacked the reduced {\bf science} images using the {\em Swarp} tool, included in the {\em Scisoft} 7.7 suite\footnote{http://www.eso.org/sci/software/scisoft}, applying a $3 \sigma$ filter to remove the hot/cold pixels and cosmic ray hits from the average image.  Since all nights were photometric, we computed the photometry calibration of the average image by using  the  average of the $R_{\rm SPEC}$  night zero points  ($< 0.05$ magnitudes rms) and extinction coefficients computed over the four nights and the colour term for the September 2010--March 2011 semester.
All these values are computed by  the  \fors\ pipeline and are available through  the \fors\  data-quality  control database\footnote{www.eso.org/qc}. For the average science image we assumed the average airmass computed over all the exposures. We estimated that the effects of the airmass variation over the different exposures (0.11 rms) and the variation of the extinction coefficient (0.012 rms) over the four nights introduce  an uncertainty of only 0.05 magnitudes on our photometry. \\ 
The astrometric solution of the FORS2 frame is based on the coordinates of the guide star used for the pointing. In order to improve the accuracy of the pulsar position on the frame, we recomputed the astrometric solution using the Two Micron All Sky Survey (2MASS) All-Sky Catalog of Point Sources (Skrutskie et al. 2006). Since the brightest stars in the field are saturated, we selected a subsample of 32 fainter, non-saturated reference stars, by excluding those at clearly bad position, e.g. close to the CCD edges. We then measured the star centroids through Gaussian fitting using the Graphical Astronomy and Image Analysis (GAIA) tool\footnote{http://star-www.dur.ac.uk/~pdraper/gaia/gaia.html}. We also used GAIA astrometric fitting routines to compute the pixel-to-sky coordinate transformation, which also account for the CCD distortions. 
 The rms of the astrometric fits was $\sigma_{r}\sim0\farcs15$ in the radial direction. To this value we added in quadrature the uncertainty $\sigma_{tr}$=0\farcs1 of the registration of the FORS2 image on the chosen astrometric reference frame. According to Lattanzi et al.\ (1997), this is  $\sigma_{tr}$=$\sqrt{n/N_{S}}\sigma_{2MASS}$, where $N_{S}$=32 is the number of stars used to compute the astrometric solution, $n$=9 is the number of free parameters in the sky--to--image transformation model, and $\sigma_{2MASS}$ is the $1 \sigma$ mean positional error of 2MASS, which depends on the brightness of the reference stars. The selected 2MASS reference stars have magnitudes $15.5 \le K \le 13$,  which corresponds to $\sigma_{2MASS} \la$0\farcs2 (Skrutskie et al.\ 2006).  We also accounted for the 0\farcs015 accuracy of the link of the 2MASS coordinates to the International Celestial Reference Frame (Skrutskie et al.\ 2006).  Thus, we estimated that the overall 1$\sigma$ positional uncertainty of our FORS2 astrometry is $\sim$0\farcs18.  For consistency, we also recomputed the astrometry calibration of the \fors\ V-band image of Mignani et al.\ (2011) using the 2MASS catalog. The resulting rms of the astrometric fits was $\sigma_{r}\sim$0\farcs16, leading to an overall 1$\sigma$ uncertainty on the V-band image of $\sim$0\farcs19.

\begin{figure}
\centering
\includegraphics[height=8cm]{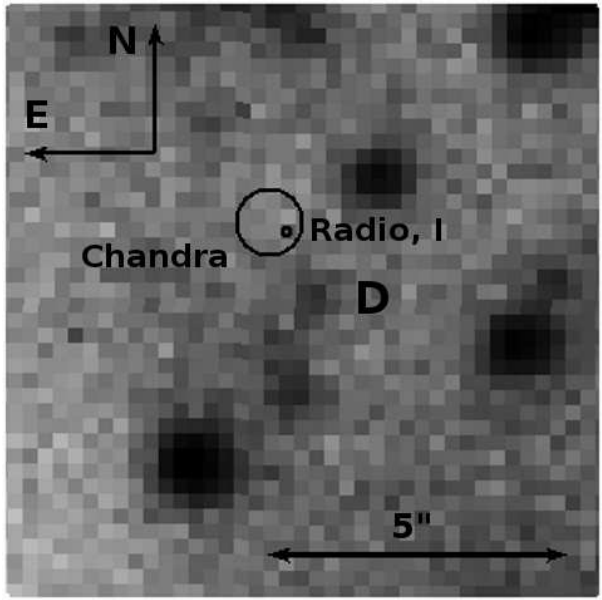}
  \caption{\vlt/\fors\  image of the \psr\ field (R band; 13797.7 s). The \chan\ (Gonzalez et al.\ 2006) and radio interferometry (Stappers et al.\ 1999)
   positions are shown as circles. The  star closest to the \chan\ position is Star D of Mignani et al.\ (2011).}
   \label{vlt}
\end{figure}

\section{Data analysis and results}

As done in Mignani et al. (2011), also in this work, we assumed as a reference both the \chan\ ($\alpha$=$10^{\rm h}  48^{\rm m} 12\fs64$; $\delta$=$-58^\circ 32\arcmin 03\farcs60$ (MJD=  52859; Gonzalez et al.\ 2006) and radio interferometry ($\alpha$=$10^{\rm h}  48^{\rm m} 12\fs604$; $\delta$=$-58^\circ 32\arcmin 03\farcs75$ (MJD=50581; Stappers et al.\  1999) positions  that have radial positional uncertainties of 0\farcs55 and  0\farcs08, respectively. 
These are shown in Fig.\ref{vlt}, overlaid on the co-added \fors\ R-band image (MJD=55566).
Our astrometry is consistent with that of Mignani et al.\ (2011), although the accuracy on our revised \fors\ astrometry is better due to the better positional accuracy of 2MASS with respect to the GSC-2 (Lasker et al.\ 2008). As seen, there is no apparent candidate counterpart to \psr\ at the expected position. The object closest to the \chan\ position is Star D of  Mignani et al.\  (2011).   We measured its flux  through standard aperture photometry, and obtained $R$=25.8$\pm$0.2, where the associated error accounts for statistical errors and the accuracy of the absolute \fors\ flux calibration (Moheler et al.\ 2010). No other star is detected at the \chan\ position down to $R\sim$26.4 ($3 \sigma$).
We ignore whether Star D was the potential pulsar counterpart claimed by Solleman \& Shibanov\footnote{http://archive.eso.org/wdb/wdb/eso/abstract/query?\&progid=386.D-0585(A)} on the basis of the \fors\ V-band data.
Nonetheless, since Star D is also detected in the new \fors\ R-band images, we used the whole data set  to  determine whether or not it can be a candidate counterpart to \psr. 

\section{Discussion}

Star D ($V$=26.7$\pm$0.2) was ruled out by Mignani et al.\ (2011) as a possible  counterpart to \psr\ because of its $\sim 3 \sigma$ offset  from the most recent \chan\ position of Gonzalez et al.\  (2006) and an $\sim$8\% chance coincidence probability.   However, since the time span between the  \chan\   (MJD=52859) and the \fors\ $V$-band  (MJD=55237) observations is $\sim$6.5 years,  the actual pulsar position could have been closer to that of Star D  at the latter epoch, owing to its unknown proper motion. For instance, for a pulsar transverse velocity of 1200 km s$^{-1}$, corresponding to a proper motion of $\sim 0\farcs9$ yr$^{-1}$ at a 2.7 kpc distance, the \psr\ position could have been closer to that of Star D by $\sim$0\farcs6.
Such a velocity is about three times as large as the average transverse velocity of radio pulsars  (Hobbs et al.\ 2005). Pulsars with extreme velocities do exist, though, such as   the Guitar Nebula pulsar, PSR\, B2224+65, whose transverse velocity could be as high as 1600 km s$^{-1}$ (Chatterjee \& Cordes 2004).  Although the probability that \psr\ is moving at such an high velocity  and exactly along a position angle of $\sim$210$^{\circ}$, i.e. that of the vector connecting the \chan\ position to that of Star D,  is small, we cannot  rule it out a priori.   {\em If} Star D was the pulsar counterpart, it should then have the same proper motion as hypothesised for the pulsar. We used the \fors\ $V$ and $R$-band images, separated by $\sim 0.9$ years, to measure a possible angular displacement of Star D through relative astrometry. We registered the two images using standard tools in {\sc IRAF}, with an rms accuracy of $\sim$ 0.05 pixel per coordinate. Unfortunately, the faintness of Star D prevented us to fits its centroid with an accuracy better than 0.2 pixel per coordinate and we could only set a $1 \sigma$ upper limit of $\sim 0.4$ pixel on its displacement. Accounting for the FORS2 pixel scale (0\farcs25), this corresponds to an upper limit of $\sim 0\farcs11$ yr$^{-1}$ on the Star D proper motion.  {\em If} Star D was the pulsar counterpart, this would only allow us to rule out a transverse velocity  $\ga 1500$ km s$^{-1}$.

We used the colour information on Star D, now available,  to further investigate its possible association with \psr. Star D has an observed $(V-R)$=0.9$\pm$0.3. 
We checked whether 
this colour would be compatible with a pulsar spectrum. We corrected the observed fluxes of Star D assuming the reddening towards \psr, computed from the hydrogen column density  inferred from the X-ray spectral fits.  
The  $N_{\rm H}$ derived from the best fit to the  \chan\ plus \xmm\ X-ray spectrum (Marelli 2012) corresponds to a reddening $E(B-V)$=0.82$\pm$0.04 according to the relation of Predehl \& Schmitt\ (1995). This is much lower than the value inferred from the \xmm\ spectrum alone (Marelli et al.\ 2011), $E(B-V)$=$1.6^{+0.7}_{-0.4}$, and assumed by Mignani et al.\ (2011) in their analysis of the \vlt\ data.  Using the interstellar extinction coefficients of Fitzpatrick (1999), we derived   $A_V$=2.55$\pm$0.13 and $A_R$=1.91$\pm$0.09.  Thus, if Star D were affected by the same extinction as the pulsar, its intrinsic colour would be  $(V-R)_0$=0.26$\pm0.33$, i.e. its spectrum would be  unconstrained.  We note that the Vela pulsar has $(V-R)_0\approx$-0.3 (Mignani \& Caraveo 2001), where the uncertainty due to the reddening correction for $N_{\rm H}=(2.2\pm0.5) \times 10^{20}$ cm$^{-2}$ (Marelli et al.\ 2011) is marginal, i.e. bluer than it would be for \psr.
However, Vela has a flat PL spectrum in the optical (Mignani et al.\ 2007), whereas  some pulsars  might have steeper spectra that rise towards longer wavelengths, such as PSR\, B0540$-$69 (Mignani et al.\ 2010a), which  would yield colours redder than Vela. Thus, the colour of Star D might not be incompatible with that of a pulsar.

We note that the flux of Star D would be incompatible with the expected optical emission of \psr\ extrapolated from the luminosity of the Vela pulsar.  \psr\ has a spin-down luminosity $\dot{E}_{SD}\sim2 \times 10^{36}$ erg s$^{-1}$,  a factor of 3 smaller than the Vela pulsar ($\dot{E}_{SD}\sim6.9 \times 10^{36}$ erg s$^{-1}$). Moreover, it is at a larger distance (2.7$\pm$0.35 kpc) than Vela (0.287$\pm$0.02 kpc; Dodson et al.\ 2003) and is affected by a factor of  $\sim$20 larger reddening.  Assuming that the \psr\  optical luminosity $L_{opt}$ is the same fraction  of the spin-down luminosity $\dot{E}_{SD}$ as the Vela pulsar and accounting for the distance and reddening would then  yield $V\approx$32.2.
If Star D were the  \psr\ counterpart, this would emit in the optical up to $\approx3.7 \times 10^{-7}$ of its spin-down luminosity. 
 This value is $\approx$100 larger than the Vela pulsar, and comparable to that of middle-aged and older pulsars (Zharikov et al.\ 2006).  However, the evolution of the pulsar optical emission efficiency, $\eta_{opt} \equiv L_{opt}/\dot{E}_{SD}$,  with the characteristic age is quite uncertain. In particular, this is true for spin-down ages between 10 and 100 krs, where Vela is the only pulsar identified in the optical (Mignani 2011). Thus, we do not know whether the sharp decrease of  $\eta_{opt}$ observed for Vela, which is only $\approx$5 times older than the Crab but has a $\approx$1000 times lower efficiency,   is peculiar of this object or is representative of  all Vela-like pulsars.  Indeed, the upper limits on the optical luminosity inferred for other Vela-like pulsars (e.g.,  Mignani et al.\ 1999; 2011; 2012)  imply emission efficiencies  $\eta_{opt} \la  10^{-6}$,  which would be consistent with that derived for \psr\ {\em if} Star D was its optical counterpart.

We investigated this association from the corresponding X-ray-to-optical flux ratio $F_{\rm X}/F_V$, where $F_{X}$ and $F_V$ are the unabsorbed \chan\ and VLT fluxes of \psr\ and Star D, respectively. For the assumed reddening, the association with Star D would imply, at most, an unabsorbed optical flux $F_V\sim8.6 \times 10^{-16}$ erg cm$^{-2}$ s$^{-1}$ and $F_{\rm X}/F_V\sim$100. 
This value 
is much lower than expected  from the comparison with other young rotation-powered pulsars (e.g., Zharikov et al. 2006), for which the non-thermal, unabsorbed X-ray-to-optical flux ratio is always larger than $\approx 800$. Thus, we conclude that Star D cannot be the \psr\ counterpart. The R-band upper limit implies a flux $F^{R}_{\nu}\la$0.46 $\mu$Jy,  corrected for the interstellar extinction.
This flux value is quite close to that derived in the V band  ($F^{V}_{\nu}\la$0.34 $\mu$Jy).  The optical flux upper limit on \psr\ imply, for the revised value of the interstellar absorption, an optical emission efficiency $\eta_{opt}\la1.9 \times 10^{-7}$, accounting for the uncertainty on the $N_{\rm H}$ and the pulsar distance.

\begin{figure}
\includegraphics[height=6.5cm,angle=0,clip=]{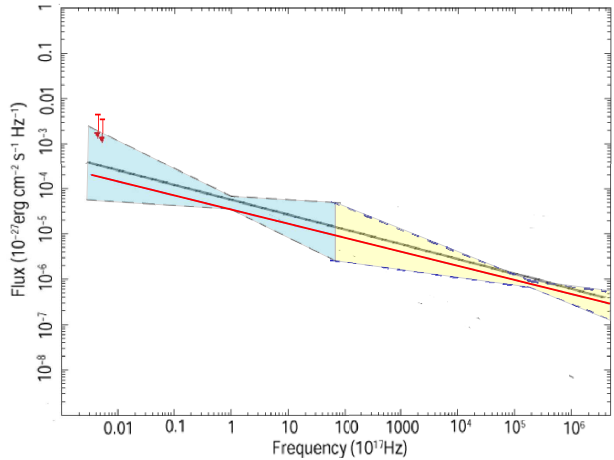}
\caption{Dereddened optical flux upper limits  of PSR\, J1048$-$5832 compared with the low-energy extrapolations of the X and $\gamma$-ray PLs  that best fit the \chan\ plus \xmm\ (Marelli 2012) and {\em Fermi}/LAT (Abdo et al.\ 2009) data (thick black and red solid lines, respectively).  The light-blue and yellow-shaded areas show the $1 \sigma$ uncertainty on the extrapolation of the X and $\gamma$-ray spectra, respectively. The uncertainty on the optical flux upper limits due to the uncertainty on the reddening is $\pm 0.04 \mu$Jy ($\pm 0.0004 \times 10^{-27}$ erg cm$^{-2}$ s$^{-1}$ Hz$^{-1}$).}
\label{sed}
 \end{figure}
 
We compared the optical flux upper limits with the low-energy extrapolation of the X-ray and $\gamma$-ray spectra (Fig.\ref{sed}). For the X-rays, we assumed the spectral model that fits the   \chan\ plus \xmm\ data (Marelli 2012),  a PL with photon index $\Gamma_{\rm X}$=1.35$\pm$0.45, while for the $\gamma$-rays we assumed the PL with photon index $\Gamma_{\gamma}$=1.38$\pm$0.13 and exponential cut-off at $\sim$2.3 GeV that fits the {\em Fermi}/LAT data (Abdo et al.\ 2009). As seen from Fig.\ref{sed}, the extrapolation of the $\gamma$-ray PL can be consistent with the X-ray PL spectrum.  This changes the conclusions reported in Mignani et al.\ (2011), based on the \xmm\ spectrum, where  the presence of a spectral break between the X-rays and the $\gamma$-rays was apparent. Fig.\ref{sed} also shows that the optical flux upper limits are just above the extrapolation of the X-ray and $\gamma$-ray spectra.  This means that the multi-wavelength spectrum of \psr\ could be consistent with a single PL stretching from the $\gamma$-rays to the optical (see also Durant et al.\ 2011), at variance with  other {\em Fermi} pulsars  (e.g., Mignani et al.\ 2010b; 2011; 2012), where there is evidence for spectral breaks between  the PL components in the different energy bands, which are possibly representative of different energy and particle distributions.  Thus, the detection of \psr\ in the optical would be crucial to confirm this scenario.

\section{Conclusions}

The new VLT observations of \psr\ confirm that the object detected closest to the \chan\ position (Star D) is  an unrelated field star. No other possible counterpart
is detected down to $R\sim$26.4 and $V\sim$27.6.  Thus, \psr\ remains unidentified in the optical.  It is clear that its detection  is close to the sensitivity limit of 8m-class telescopes. Observations at longer wavelength might be more successful, though, were the optical/infrared spectrum of \psr\  follow the extrapolation of the X-ray PL.  An updated \chan\ or radio position for \psr\ would allow one to minimise the position uncertainty due to the unknown pulsar proper motion and would represent an advantage in the search for its optical/infrared counterpart.

\section*{Acknowledgments} We thank the anonymous referee whose comments has contributed to improve the quality of our manuscript.

\label{lastpage}

\end{document}